\documentclass[conference]{IEEEtran}
\usepackage{cite}
\usepackage{amsmath,amssymb,amsfonts,amsthm}
\usepackage{algorithmic}
\usepackage{graphicx}
\usepackage{textcomp}
\usepackage{xcolor}
\usepackage{float}
\usepackage{afterpage}
\usepackage{bm}
\usepackage{bbold}
\usepackage{tcolorbox}
\usepackage{diagbox}
\usepackage{tabularx}
\usepackage{subcaption}
\usepackage[normalem]{ulem}

\usepackage{pgfplotstable}
\usepgfplotslibrary{fillbetween}

\newtheorem{definition}{Definition}

\usepackage{threeparttable}
\usepackage[linesnumbered,ruled,vlined]{algorithm2e}
\usepackage[absolute,overlay]{textpos}
\usepackage{tikz}
\usetikzlibrary{shapes.geometric, arrows.meta, positioning}
\usetikzlibrary[datavisualization]

\graphicspath{ {./figures/}}
\def\BibTeX{{\rm B\kern-.05em{\sc i\kern-.025em b}\kern-.08em
    T\kern-.1667em\lower.7ex\hbox{E}\kern-.125emX}}

\tikzstyle{startstop} = [ellipse, minimum width=1cm, minimum height=0.5cm,text width=4.5em, text width=4.5em, text centered, draw=black]
\tikzstyle{decision} = [diamond, draw, fill=white, text width=9em, text centered, node distance=2cm, inner sep=0pt]
\tikzstyle{block} = [rectangle, draw, fill=white, text width=22em, inner sep=1em, rounded corners, minimum height=4em]
\tikzstyle{blockprocess} = [rectangle, draw, fill=white, text width=12em, text centered, inner sep=1em, rounded corners, minimum height=4em]
\tikzstyle{line} = [draw, -{Latex[length=2mm]}]
\tikzstyle{arrow} = [thick,->,>=stealth]
\tikzstyle{output} = [trapezium, trapezium left angle=70, trapezium right angle=110, minimum width=1cm, minimum height=0.5cm, text centered, draw=black]

\newcommand{\wrt}{\textit{w.r.t.~}}

\newcommand{\ie}{\textit{i.e.~}}

\newcommand{\eg}{\textit{e.g.~}}

\definecolor{lightgray}{gray}{0.35}

\title{Early Acceptance Matching Game for User-Centric Clustering in Scalable Cell-free MIMO Networks\thanks{This work has been accepted for publication in 2024 European Conference on Networks and Communications (EuCNC) \& 6G Summit}} 
\author{Ala Eddine Nouali\IEEEauthorrefmark{1}, Mohamed Sana\IEEEauthorrefmark{1}, Jean-Paul Jamont\IEEEauthorrefmark{2}
\smallskip\\
\IEEEauthorrefmark{1}CEA-Leti, Université Grenoble Alpes, F-38000 Grenoble, France,\\
\IEEEauthorrefmark{2}LCIS, Université Grenoble Alpes, Valence, France\\
\IEEEauthorblockA{
Email : \{ala-eddine.nouali, mohamed.sana\}@cea.fr; jean-paul.jamont@univ-grenoble-alpes.fr
}}

\newcommand{\titleheader}{This work has been accepted for publication in 2024 European Conference on Networks and Communications (EuCNC) \& 6G Summit}
\def\ps@IEEEtitlepagestyle{%
\def\@oddhead{\mbox{}\scriptsize \titleheader \rightmark \hfil}%
}
\makeatother

\begin{document}
\maketitle
\begin{abstract}
The canonical setup is the primary approach adopted in cell-free multiple-input multiple-output (MIMO) networks, in which all access points (APs) jointly serve every user equipment (UE). This approach is not scalable in terms of computational complexity and fronthaul signaling becoming impractical in large networks. This work adopts a user-centric approach, a scalable alternative in which only a set of preferred APs jointly serve a UE. Forming the optimal cluster of APs for each UE is a challenging task, especially, when it needs to be dynamically adjusted to meet the quality of service (QoS) requirements of the UE. This complexity is even exacerbated when considering the constrained fronthaul capacity of the UE and the AP. We solve this problem with a novel many-to-many matching game. More specifically, we devise an early acceptance matching algorithm, which immediately admits or rejects UEs based on their requests and  available radio resources. The proposed solution significantly reduces the fronthaul signaling while satisfying the maximum of UEs in terms of requested QoS compared to state-of-the-art approaches.
\end{abstract}

\section{Introduction}
Commercialized networks use cellular architecture, where terminals suffer from large variation of quality of service (QoS) \cite{Foundations_Cell_free_Massive_MIMO}. User equipment (UE) close to the access point (AP), which references the cell center can achieve a high QoS. However, the QoS fluctuates significantly as user moves further away from the center, preventing cellular networks from providing high, consistent and uniform QoS to every UE regardless of their location in the network. Cell-free aims to solve this problem. It is a proposed new
paradigm that eliminates cell boundaries and manages interference by joint processing at APs to serve UEs \cite{Cell_free_massive_MIMO_versus_small_cells}. The canonical setup is the trivial approach adopted in cell-free networks, in which all APs jointly serve each UE. 
This approach does not scale well in terms of fronthaul requirements and computational complexity, becoming impractical in large networks \cite{Scalable_cell_free_massive_MIMO_systems}. User-centric clustering is the proposed scalable alternative, in which each UE is served by the nearest preferred APs \cite{Cell_free_massive_MIMO_User_centric_approach, User_centric_communications_versus_cell_free_massive_MIMO_for_5G_cellular_networks}. 
\begin{figure}[t]
    \centering
    \includegraphics[width=\columnwidth]{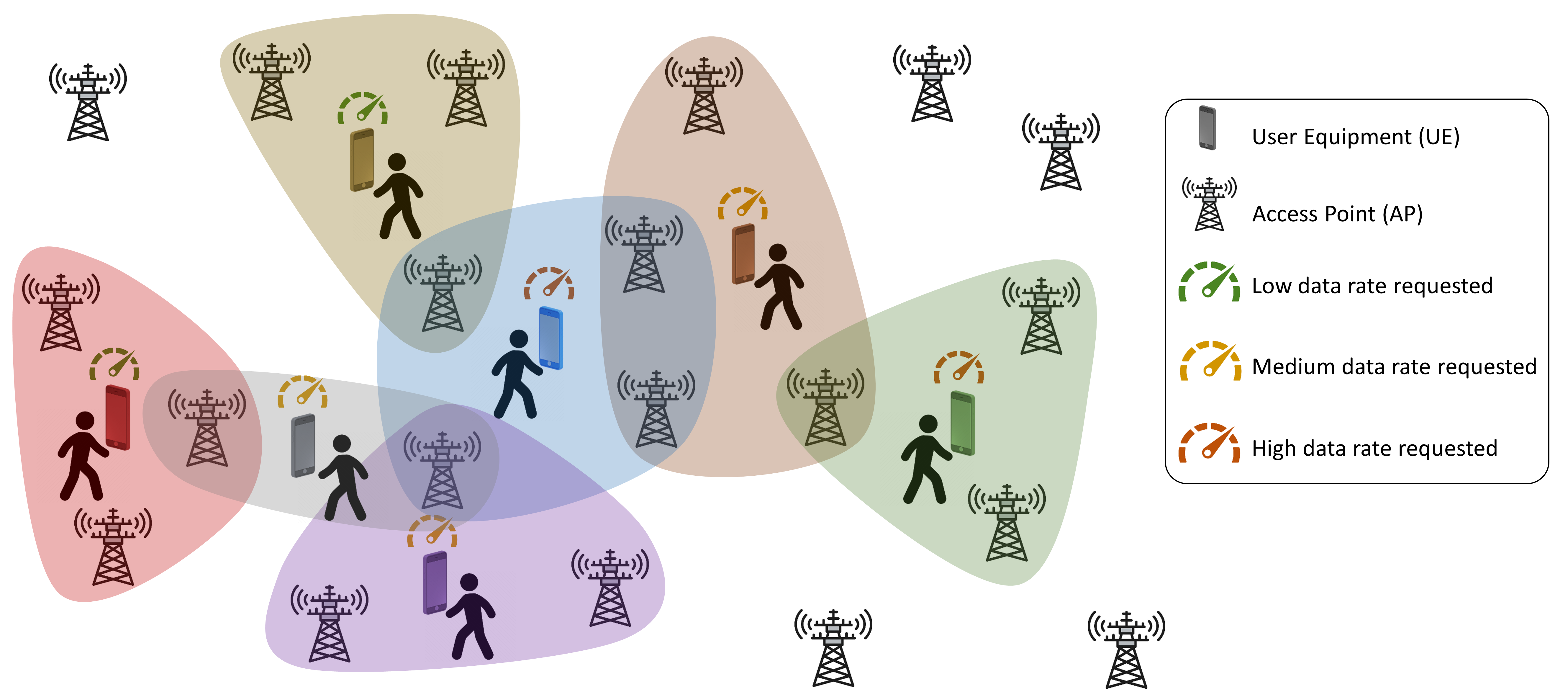}
    \caption{A user-centric clustering operation in a cell-free MIMO network. Here, $M$ geographically distributed APs jointly serve $K$ UEs. Each UE is served by a cluster of nearby preferred APs depending on its QoS requirements.}
    \label{fig:user-centric clustering}
\end{figure}
Different techniques are adopted to form clusters of APs for UEs. The works \cite{Scalable_cell_free_massive_MIMO_systems, Scalable_pilot_assignment_for_user_centric_cell_free_massive_MIMO_networks} study centralized and 
distributed association AP-UE based on pilot assignment. Authors in
\cite{Cluster_formation_in_scalable_cell_free_massive_MIMO_networks,User_Centric_Clustering_in_Cell_Free_MIMO_Networks_using_Deep_Reinforcement_Learning} propose to minimize fronthaul signaling while guaranteeing minimum QoS for UEs.
Due to limited network resource, which need to be efficiently allocated according to demand and network conditions, various studies investigate user-centric clustering to manage resource allocation in cell-free networks. For instance, \cite{Joint_AP_On_Off_and_User_Centric_Clustering_for_Energy_Efficient_Cell_Free_Massive_MIMO_Systems} maximizes network energy efficiency constrained to UEs QoS and APs transmission power capabilities. 
Similarly, \cite{Downlink_resource_allocation_in_multiuser_cell_free_MIMO_networks_with_user_centric_clustering, Whale_Swarm_Reinforcement_Learning_Based_Dynamic_Cooperation_Clustering_Method_for_Cell_Free_Massive_MIMO_Systems} study cluster formation to maximize the network throughput.
Although there are works that restrict the total number of APs serving a UE, these works consider that the capacity of the APs is unlimited: an AP can accept any number of UEs. This assumption is unrealistic as in practice, UEs and APs have limited fronthaul capacity and computational capability. Considering these constraints, various methods have been employed to allocate radio resources efficiently, with many-to-many (M2M) matching game emerging as a particularly effective solution \cite{Matching_theory_for_future_wireless_networks_Fundamentals_and_applications}. This approach is based on the principles of matching theory, which has attracted the interest of researchers because of its ability to enable distributed user association with low computational complexity and fast convergence time \cite{Matching_theory_for_future_wireless_networks_Fundamentals_and_applications}. 
Deferred acceptance (DA) is the common strategy used in M2M matching game. In this method, APs buffer user association requests at each round, finalizing the association procedure only at the last iteration \cite{Many_to_Many_matching_user_association_scheme_in_ultra_dense_millimeter_wave_networks}. This results in a prohibitive delays and slow convergence time.

Different from previous work, we propose to optimize radio resource allocation by forming clusters of APs for UEs depending on their specific QoS requirements (as shown in Fig. \ref{fig:user-centric clustering}). We present a novel user association scheme based on M2M matching game between the set of UEs and APs, aiming to satisfy the maximum UEs in the network in terms of requested QoS. In contrast to the DA strategy proposed in \cite{Many_to_Many_matching_user_association_scheme_in_ultra_dense_millimeter_wave_networks}, we adopt an early acceptance (EA) strategy to speed up the AP-UE association process. In this approach, UEs apply to a specific set of APs, which immediately accept or reject the association requests based on their preferences and available radio resources.
The proposed solution significantly reduces the computational complexity of the cluster formation process, while providing an acceptable satisfaction level for the maximum number of UEs in the network, depending on their requested QoS.


\section{System model \& problem formulation} \label{sec:sys-model}
\subsection{System Model}

We consider a downlink cell-free MIMO network consists of $M$ geographically distributed APs that cooperate to serve $K$ geographically distributed UEs. Let $\mathcal{M} = \{1,\dots,M\}$ and $\mathcal{K} = \{1,\dots,K\}$ denote the set of APs and UEs, respectively. Each AP $m$ is equipped with $N$ antennas, while each UE has a single antenna. 
We consider the Rayleigh fading channel model adopted in \cite{Cluster_formation_in_scalable_cell_free_massive_MIMO_networks}, in which the complex random channel $\mathbf{h}_{k,m} \in \mathcal{C}^{N \times 1}$, independent and identically distributed (iid), between UE $k$ and AP $m$ at time $t$ reads as:
\begin{equation}
    \label{h_{k,m}}
    \mathbf{h}_{k,m}(t) = \mathbf{\alpha}_{k,m}(t)\sqrt{g_{k,m}(t)},
\end{equation}
where $\mathbf{\alpha}_{k,m}(t) \sim \mathcal{CN}(0, I_{N})$ denotes the small-scale fading and $g_{k,m}(t)$ is the channel gain, which captures the distance-dependent path loss and the shadowing effect as follows:
\begin{equation}
    g_{k,m}(t) = \left(\frac{\lambda}{4\pi}\right)^2\left(\frac{1}{d_{k, m}(t)}\right)^{\eta}\chi_{k, m}(t).
\end{equation}
Here, $\lambda$ is the wavelength of the operated carrier frequency $f$, $d_{k, m}$ is the distance between UE $k$ and AP $m$, $\eta$ denotes the path loss exponent and $\chi_{k,m} \sim \mathcal{LN}(0, \sigma_{s}^{2})$ denotes the random log-normally distributed shadowing. 

In the downlink scenario, we assume a full knowledge of channels between APs and UEs. We denote with $\mathbf{v}_{k,m}(t) \in \mathcal{C}^{N \times 1}$ the downlink transmit beamformer at AP $m$ for UE $k$ at time $t$, which we obtain via the linear minimum mean squared error (LMMSE) precoding algorithm \cite{LMMSE}:
\begin{equation}
    \mathbf{v}_{k,m}(t) = \mathbf{h}_{k,m}(t)(\mathbf{h}_{k,m}^H(t)\mathbf{h}_{k,m}(t) + \sigma_n^2(t))^{-1}.
\end{equation}
Here, $\mathbf{h}_{k,m}^H$ is the conjugate transpose of $\mathbf{h}_{k,m}$ and $\sigma_{n}^2(t)$ denotes the receiver noise power at time $t$, which is a scalar in our case. Accordingly, the
signal-to-interference-plus-noise ratio (SINR) perceived by UE $k$ at time $t$ reads as:
\begin{equation}
    \mathrm{SINR}_k(t) = \frac{S_k(t)}{I_k(t)+\sigma_n^2(t)},
\end{equation}
where $S_k(t)$ is the received signal power given as:
\begin{equation} \label{eq:received_power}
    S_k(t) = \left|\displaystyle\sum_{m=1}^M \sqrt{P_{k, m}(t)} \mathbf{h}_{k, m}^H(t) \mathbf{v}_{k, m}(t) \delta_{k, m}(t)\right|^2,
\end{equation}
and $I_k(t)$ is the perceived interference power given as:
\begin{equation}\label{eq:interference}
    I_k(t) = \displaystyle\sum_{\substack{j=1 \\ j \neq k}}^K\left|\displaystyle\sum_{m=1}^M \sqrt{P_{j, m}(t)} \mathbf{h}_{k, m}^H(t) \mathbf{v}_{j, m}(t) \delta_{j, m}(t)\right|^2.
\end{equation}
In Eq. \eqref{eq:received_power} and \eqref{eq:interference}, $P_{k,m}(t)$ denotes the power allocated by AP $m$ to UE $k$ at time $t$ and $\delta_{k,m}(t)$ is a binary association variable, which indicates 
whether UE $k$ is associated with AP $m$ at time $t$, in which case $\delta_{k,m}(t) =1 $ and $\delta_{k,m}(t) = 0$ otherwise.
Hence, the data rate perceived by UE $k$ at time $t$ is given by:
\begin{equation}
    R_{k}(t) = B \cdot \mathrm{SE}_{k}(t),
\end{equation}
where $\mathrm{SE}_{k}(t) = \log_{2}(1 + \mathrm{SINR}_{k}(t))$ is the spectral efficiency (SE) of UE $k$ at time $t$ and $B$ is the total system bandwidth. 
\subsection{Problem Formulation}

Let $R_{k}^{\rm req}(t)$ denote the time-varying data rate demand of UE $k$ and $\kappa_k(t)$ indicates its QoS satisfaction 
at time $t$, which we define as in \cite{sana2021transferable} as follows:
\begin{equation}\label{eq:QoS}
    \kappa_k(t) = \min\left(1, \frac{R_{k}(t)}{R_{k}^{\mathrm{req}}(t)}\right).
\end{equation}
Let $\kappa_0\in[0,1]$ define the minimum QoS satisfaction level of UEs. We say a UE $k$ is $\kappa_0$-satisfied if $\kappa_k(t) \geq \kappa_0$. In this case, when $\kappa_0=1$, we say the UE is fully satisfied. The QoS of a UE varies according to multiple factors, including power allocation and user-clustering strategy, which affect cell interference and hence network performance.

In this context, we are interested in maximizing the total number of $\kappa_0$-satisfied UEs, while minimizing the number of fronthaul associations. To this end, we propose the following optimization problem:
\begin{align}
\underset{\Psi(t)}{\max} &  \quad \mathbb{E}\left\{\displaystyle\sum_{\substack{k \in \mathcal{K}}} \mathbb{1}_{\{\kappa_{k}(t) \geq \kappa_{0}\}}\right\}, \tag{$\mathcal{P}$} \label{eq:P}\\[0cm]
    	\mathrm{s.t.~}~ & {} \delta_{k,m}(t) \in \{0,1\},  \qquad \forall k\in \mathcal{K},m\in \mathcal{M}, &\tag{$\mathcal{C}_{1}$} \label{C1}\\
	    {}&P_{k,m}(t) \geq 0, \qquad \forall k\in \mathcal{K},m\in \mathcal{M}, & \tag{$\mathcal{C}_{2}$} \label{C2}\\
		{}&\displaystyle\sum_{\substack{k \in \mathcal{K}}} \delta_{k,m}(t) P_{k,m}(t)  \leq P_{\max}, \qquad \forall m \in \mathcal{M}, &\tag{$\mathcal{C}_{3}$} \label{C3}\\
        {}&\sum_{\substack{k \in \mathcal{K}}} \delta_{k,m}(t) \leq K_{\max}, \qquad   \forall m \in \mathcal{M}, & \tag{$\mathcal{C}_{4}$} \label{C4}\\
        {}&\sum_{\substack{m \in \mathcal{M}}} \delta_{k,m}(t) \leq M_{\max}, \qquad  \forall k \in \mathcal{K},   \tag{$\mathcal{C}_{5}$} \label{C5}
\end{align}
where $\Psi(t) = \{\delta_{k,m}(t),P_{k,m}(t),\;\forall k,m \in \mathcal{K}\times\mathcal{M}\}$ is the set of parameters to adjust to maximize our objective function. 
The expectation in \eqref{eq:P} is taken \wrt the random traffic requests and channel realizations. The constraint \eqref{C1} defines the variable of association $\delta_{k,m}$
as a binary variable. Constraints \eqref{C2} and \eqref{C3} restricts the transmission power of an AP $m$ to do not exceed its maximum of transmission power $P_{\max}$. Constraint \eqref{C4} indicates that each AP $m$ can
serve $K_{\max}$ UEs at most. Finally, constraint \eqref{C5} ensures that each UE $k$ is associated to at most $M_{\max}$ APs.

In this work, we focus on the user-clustering problem. We assume that each AP equally shares its transmit power with its served UEs. Thus, the power allocation strategy is fully determined by the user-clustering operation.


\section{User-centric clustering as a many-to-many matching game}\label{sec:game}

\begin{figure}[t]
    \centering
    \resizebox{0.8\columnwidth}{!}{
    \begin{textblock*}{2cm}(1.34cm,1.775cm)
        \includegraphics[width=0.2\textwidth]{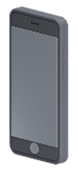}
    \end{textblock*}

    \begin{textblock*}{2cm}(1.34cm,3.7cm)
        \includegraphics[width=0.4\textwidth]{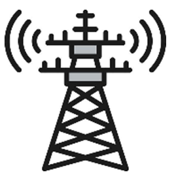}
    \end{textblock*}

    \begin{tikzpicture}[sbs/.style={draw, circle, fill=orange!30, minimum size=1.5cm},
                        ue/.style={draw, circle, fill=green!30, minimum size=1.5cm},
                        set/.style={draw, rectangle, fill=blue!20, minimum width=3cm, minimum height=2cm}]

    \node[set, label=left:\huge{AP set}, minimum width=13cm] (sbsset) at (5,-3) {};
    \node[sbs] (sbs1) at (0,-3) {AP $1$};
    \node[sbs] (sbs2) at (2,-3) {AP $2$};
    \node[sbs] (sbsd1) at (4,-3) {$\dots$};
    \node[sbs] (sbsn) at (6,-3) {AP $m$};
    \node[sbs] (sbsd2) at (8,-3) {$\dots$};
    \node[sbs] (sbsN) at (10,-3) {AP $M$};

    \node[set, label=left:\huge{UE set}, minimum width=13cm] (ueset) at (5,1) {};
    \node[ue] (ue1) at (0,1) {UE $1$};
    \node[ue] (ue2) at (2,1) {UE $2$};
    \node[ue] (ued1) at (4,1) {$\dots$};
    \node[ue] (uek) at (6,1) {UE $k$};
    \node[ue] (ued2) at (8,1) {$\dots$};
    \node[ue] (ueK) at (10,1) {UE $K$};

    \draw (sbs1) -- (ue1);
    \draw (sbs1) -- (ue2);
    \draw (sbs1) -- (uek);
    \draw (sbs2) -- (ue2);
    \draw (sbs2) -- (uek);
    \draw (sbsn) -- (ue1);
    \draw (sbsn) -- (ueK);
    \draw (sbsN) -- (ue2);
    \draw (sbsN) -- (uek);
    \draw (sbsN) -- (ueK);
    \end{tikzpicture}
    }
    \caption{Many-to-many matching between APs and UEs.}
    \label{fig:many_to_many_illustration}
\end{figure}
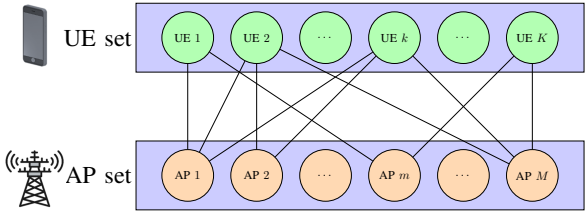

As shown in Fig. \ref{fig:many_to_many_illustration}, user association can be seen as a many-to-many matching game between two sets of players: APs and UEs.
During this game, each player, from a set of local observations, constructs a preference list based on its own objective. UEs request association to APs according to their preference lists and APs, in turn, decide individually the UEs to serve based on their own preference lists.

\subsection{Background on many-to-many matching concepts}

Before formulating the user-centric clustering as a many-to-many matching game, we first introduce some basic concepts based on two-sided matching theory \cite{Matching_theory_for_future_wireless_networks_Fundamentals_and_applications}.

In matching game, each UE $k$ starts by building a preference list of ordered APs $\mathcal{P}_{k}^{\rm UE}$, from most to least preferred. To do so, it relies on a preference metric $\vartheta_{k,m}^{\rm UE}$ (\eg the perceived SINR \wrt AP $m$): UE $k$ prefers AP $m$ to AP $m'$ ($m \neq m'$) if $\vartheta_{k,m}^{\rm UE} \geq \vartheta_{k,m'}^{\rm UE}$. Similarly, each AP $m$ uses a preference metric $\vartheta_{k,m}^{\rm AP}$ (\eg the channel gain $g_{k,m}$ \wrt UE $k$) to build its preference list of ordered UEs $\mathcal{P}_{m}^{\rm AP}$.

\begin{definition}[]
A many-to-many matching $\mu$ is a mapping function that assigns a matching vector of APs $\mathcal{C}_{k}^{\rm UE}$ to each UE $k$ and a matching vector of UEs $\mathcal{C}_{m}^{\rm AP}$ to each AP $m$. 
The matching process is constrained to:
\begin{enumerate}
    \item $\mu(k) = \mathcal{C}_{k}^{\rm UE}  \subseteq \mathcal{M}$ and $|\mu(k)| \leq M_{\max}$ $\forall k \in \mathcal{K}$,
    \item $\mu(m) = \mathcal{C}_{m}^{\rm AP} \subseteq \mathcal{K}$ and $|\mu(m)| \leq K_{\max}$ $\forall m \in \mathcal{M}$,
    \item $k \in \mu(m) \Leftrightarrow m \in \mu(k)$.
\end{enumerate}
\end{definition}
Conditions (1) and (2) represent the matching process between APs and UEs whereas condition (3) guarantees a mutually accepted match between the set of UEs and APs.

The initial step in a matching game involves constructing the preference lists of the players. In the case of wireless networks, we can build these preference lists through various metrics including channels gains, UEs SINR or data rates.
In our case, we build the preference lists for both UEs and APs based on the channel gains.

\subsection{Proposed algorithm}

\emph{Deferred acceptance} (DA) is a two-sided matching game in which, at each iteration, each AP retains in its waiting list only the $K_{\max}$ UEs preferred among the first new UE applicants and those previously in its waiting list, and reject the others \cite{Many_to_Many_matching_user_association_scheme_in_ultra_dense_millimeter_wave_networks}.
UEs in APs waiting list will be associated after the final iteration of the game \ie the association procedure is deferred until the end of the game. As a result, the association between APs and UEs can be a time-consuming process. 

To overcome this problem, we adopt a new matching game called \emph{early acceptance} (EA) with preference list updating (EA-PLU) \cite{Distributed_user_association_in_B5G_networks_using_early_acceptance_matching_game}.
In this game, APs immediately decide to accept or reject UEs at each iteration. This procedure allows to accelerate the association process and control the number of associations between APs and UEs.
Consequently, EA game reduces the total number of fronthaul connections unlike DA game with known number of associations equal to $\min(MK_{\max}, KM_{\max}$). With lower AP-UE connections, few messages are exchanged between UEs and APs via fronthaul links, thus, limiting the communication overhead. 

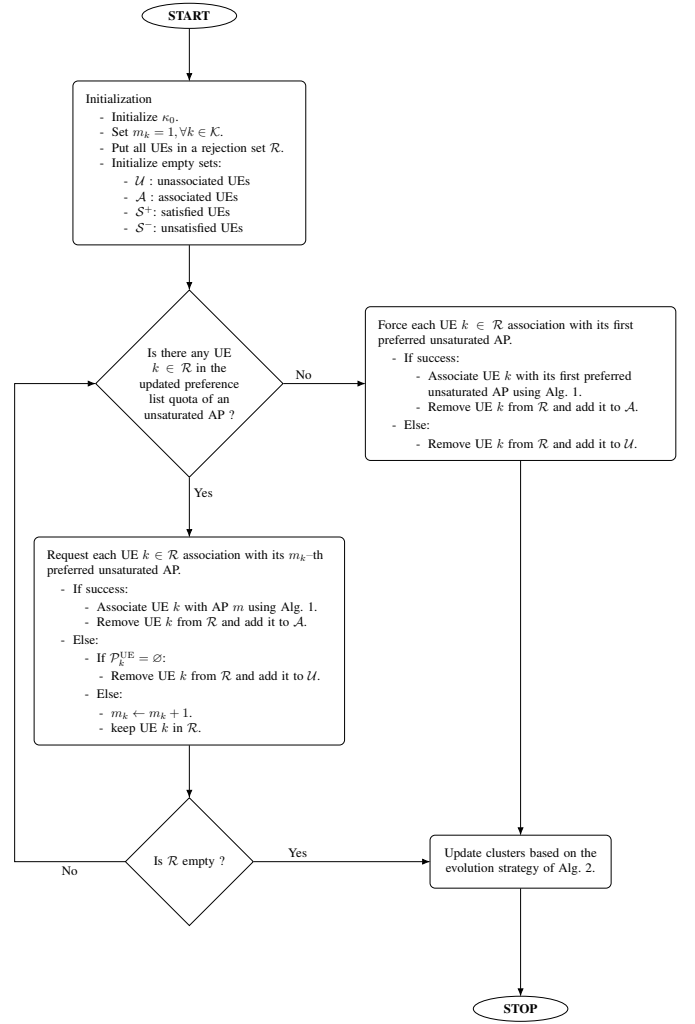
\begin{figure}[!t]
    \centering
    \resizebox{\columnwidth}{!}{
    \begin{tikzpicture}[node distance = 2cm, auto]
        \node [startstop] (start) {\textbf{START}};
        \node [block, below of=start, text width=16em, node distance=4cm] (block1) {
        Initialization
            \begin{itemize}
                \item[-] Initialize $\kappa_{0}$.
                \item[-] Set $m_{k} = 1, \forall k \in \mathcal{K}$.
                \item[-] Put all UEs in a rejection set $\mathcal{R}$.
                \item[-] Initialize empty sets:
                \begin{itemize}
                    \item[-] $\mathcal{U}~$: unassociated UEs
                    \item[-] $\mathcal{A}~$: associated UEs
                    \item[-] $\mathcal{S}^+$: satisfied UEs
                    \item[-] $\mathcal{S}^-$: unsatisfied UEs
                \end{itemize}
            \end{itemize}
            };
        \node [decision, below of=block1, node distance=6cm] (dec1) {Is there any UE $k \in \mathcal{R}$ in the updated preference list quota of an unsaturated AP ?};
        \node [block, below of=dec1,  text width=22em, node distance=7cm] (block2) {Request each UE $k \in \mathcal{R}$ association with  its $m_{k}–$th preferred unsaturated AP.
        \begin{itemize}
            \item[-] If success:
            \begin{itemize}
                \item[-]  Associate UE $k$ with AP $m$ using Alg. \ref{alg:association}.\;
                \item[-] Remove UE $k$ from $\mathcal{R}$ and add it to $\mathcal{A}$.
            \end{itemize}
            \item[-] Else:
            \begin{itemize}
               \item[-]  If $\mathcal{P}_{k}^{\rm UE} = \varnothing$:
               \begin{itemize}
                   \item[-]  Remove UE $k$ from  $\mathcal{R}$ and add it to $\mathcal{U}$.
               \end{itemize}
               \item[-]  Else:
               \begin{itemize}
                    \item[-] $m_{k} \leftarrow m_{k} + 1$.
                    \item[-] keep UE $k$ in $\mathcal{R}$.
               \end{itemize}
            \end{itemize}
        \end{itemize}
    };
        \node [block, right of=dec1, node distance=9cm] (block3) {Force each UE $k \in \mathcal{R}$ association with its first preferred unsaturated AP.
        \begin{itemize}
            \item[-] If success:
            \begin{itemize}
                \item[-] Associate UE $k$ with its first preferred unsaturated AP using Alg. \ref{alg:association}.
                \item[-]  Remove UE $k$ from $\mathcal{R}$ and add it to $\mathcal{A}$.
            \end{itemize}
            \item[-] Else:
            \begin{itemize}
                \item[-] Remove UE $k$ from $\mathcal{R}$ and add  it to $\mathcal{U}$.
            \end{itemize}
        \end{itemize}
    };
        \node [decision, below of=block2, node distance=6cm] (dec2) {Is $\mathcal{R}$ empty ?};
        \node [blockprocess, right of=dec2, node distance=9cm] (block4) {Update clusters based on the evolution strategy of Alg. \ref{alg:algorithm3}.};
        \node [startstop, below of=block4, node distance=4cm] (stop) {\textbf{STOP}};

        \path [line] (start) -- (block1);
        \path [line] (block1) -- (dec1);
        \path [line] (dec1) -- node [near start] {Yes} (block2);
        \path [line] (dec1) -- node [near start] {No} (block3);
        \path [line] (block2) -- (dec2);
        \path [line] (dec2) -- node [near start] {Yes} (block4);
        \path [line] (block3) -- (block4);
        \path [line] (dec2.west) -| node [near start] {No} ([xshift=-3cm]dec2.west)  |- (dec1.west);
        \path [line] (block4) -- (stop);
    \end{tikzpicture}}
    \caption{Flowchart representing our EA user-centric clustering procedure.}
    \label{fig:flowchart1}
\end{figure}

Our algorithm is summarized in the flowchart of Fig. \ref{fig:flowchart1}. The proposed EA game takes as an input the preference lists of APs ($\mathcal{P}_{m}^{\rm AP}, \forall m \in \mathcal{M}$) and UEs ($\mathcal{P}_{k}^{\rm UE}, \forall k \in \mathcal{K}$) and the quotas of APs ($q_{m}^{\rm AP},  \forall m \in \mathcal{M}$) and UEs ($q_{k}^{\rm UE}, \forall k \in \mathcal{K}$).
The quota of UE $k$ (resp. AP $m$) is the number of remaining possible associations it can establish. The output of the EA game is the matching
vector for UEs $\mathcal{C}^{\rm UE}=\left[\mathcal{C}^{\rm UE}_{1}, \mathcal{C}^{\rm UE}_{2}, \ldots, \mathcal{C}^{\rm UE}_{K}\right]$.

\begin{algorithm}[!b]
    \caption{User association between ${\rm AP}_{m}$ and ${\rm UE}_{k}$}
    \label{alg:association}
    Add AP $m$ to $\mathcal{C}^{\rm UE}_{k}$\;
    Remove UE $k$ from $\mathcal{P}_{m}^{\rm AP}$\;
    Remove AP $m$ from $\mathcal{P}_{k}^{\rm UE}$\;
    $q_{k}^{\rm UE} \leftarrow q_{k}^{\rm UE}-1$\;
    $q_{m}^{\rm AP} \leftarrow q_{m}^{\rm AP}-1$\;
    \If{$q_{m}^{\rm AP}=0$}{
    Remove AP $m$ from $\mathcal{P}_{k}^{\rm UE}, \forall k \in \mathcal{K}$\;}
    \If{$q_{k}^{\rm UE}=0$}{ 
        Remove UE $k$ from $\mathcal{P}_{m}^{\rm AP},  \forall m \in \mathcal{M}$\;}
    \label{alg:algorithm2}
\end{algorithm}

We initialize our algorithm by setting the preference index of all UEs to one ($m_{k}=1, \forall k \in \mathcal{K}$),
forming a rejection set $\mathcal{R}$ containing all UEs, and creating empty sets of associated UEs ($\mathcal{A}=\varnothing$),
unassociated UEs ($\mathcal{U}=\varnothing$), satisfied UEs ($\mathcal{S}^{+}=\varnothing$) and unsatisfied UEs ($\mathcal{S}^-=\varnothing$). The first step of our algorithm consists in maximizing the number of associated UEs. We start by associating each UE to one AP. At each iteration of the game, each UE $k$ applies to each $m_{k}$-th preferred AP $m$ and it will be immediately accepted if it is among the top-$q_{m}^{\rm AP}$ UEs in the preference list of the AP $m$. \textbf{Algorithm 1} details the association procedure between UE $k$ and AP $m$.

When the updated preference list of a UE $k$ becomes empty, it will be added to the set of unassociated UEs $\mathcal{U}$. UEs that are rejected by all APs remain in $\mathcal{R}$. For each UE $k$ remaining in $\mathcal{R}$ (not preferred by APs), we force the association with its first AP $m$ in its updated preference list even though it is not among the top-$q_{m}^{\rm AP}$ UEs in its updated preference list. When all APs run out of quota, UEs remaining in $\mathcal{R}$ will be added to the set of unassociated UEs $\mathcal{U}$. Through those forced associations, we provide best link quality for UEs remaining in $\mathcal{R}$ 
after prioritizing UEs favored by APs in the previous step. In this way, we guarantee that either all UEs are served by at least one AP or the quotas of the APs are fully exploited.

\begin{definition}[favorable-association pair] \label{def:def2}
   (${\rm AP}_{m}$, ${\rm UE}_{k}$) is a favorable-association pair if and only if it satisfies
\begin{enumerate}
    \item $\mathrm{UE}$ $k \in \mathcal{P}_{m}^{\rm AP}\left(1: q_{m}^{\rm AP}\right)$,
    \item Associate ${\rm UE}_{k}$ to ${\rm AP}_{m}$ $\Rightarrow$ $\kappa_{k}(\mu_{\rm evolve}) > \kappa_{k}(\mu)$\\ and  $\displaystyle\sum_{\substack{i \in \mathcal{K} \backslash \mathcal{U}}}\kappa_{i}(\mu_{\rm evolve}) \geq \displaystyle\sum_{\substack{i \in \mathcal{K} \backslash \mathcal{U}}}\kappa_{i}(\mu)$,
\end{enumerate} 
where $\mu_{\rm evolve}$ is the matching strategy obtained after letting $\mu$ evolve with the new association between AP $m$ and UE $k$.
\end{definition}

\setlength{\headsep}{0.4in}
\begin{algorithm}[!t]
    \small
    \caption{Cluster evolution process}
    \label{alg:algorithm3}
    \KwData{ $\mathcal{P}_{m}^{\rm AP}, \mathcal{P}_{k}^{\rm UE}, q_{m}^{\rm AP}, q_{k}^{\rm UE}, \forall k \in \mathcal{K}, m \in \mathcal{M}$, \\ \hspace*{1.25cm}and sets $\mathcal{A}, \mathcal{S}^+$ and $\mathcal{S}^-$}
    \KwResult{ Matching vector $\mathcal{C}^{\rm UE}$}
    \While{$\mathcal{A} \neq \varnothing$ and there exists a favorable-association pair}{
    Each UE $k \in \mathcal{A}$ tests its satisfaction\;
    \eIf{$\kappa_{k} \geq \kappa_{0}$}{
    Remove UE $k$ from $\mathcal{A}$ and add it to $\mathcal{S}^+$\;
    }{
    \eIf{$\mathcal{P}_{k}^{\rm UE} \neq \varnothing$}{
    Keep UE $k$ in $\mathcal{A}$ \;
    }{
    Remove UE $k$ from $\mathcal{A}$ and add it to $\mathcal{S}^-$\;
    }
    }
    Each UE $k \in \mathcal{A}$ tries to improve its $\kappa_{k}$\;
    \eIf{$\mathcal{P}_{k}^{\rm UE} \neq \varnothing$}{
    Set $m_{k}=1$\;
    \While{$m_{k} \leq \min(q_{k}^{\rm UE}, |\mathcal{P}_{k}^{\rm UE}|)$ and there exists no favorable-association pair}{
    UE $k$ applies to its $m_{k}$-th preferred AP (namely AP $m$ with $q_{m}^{\rm AP} \neq 0$)\;
    \eIf{$({\rm AP}_{m}, {\rm UE}_{k})$ is a favorable-association pair}{
    Associate ${\rm UE}_{k}$ with ${\rm AP}_{m}$ using Alg. \ref{alg:association}\;
    Keep UE $k$ in $\mathcal{A}$ \;
    }{
    $m_{k} \leftarrow m_{k} + 1$ \;
    }
    }
    }{
    Remove UE $k$ from $\mathcal{A}$ and add it to $\mathcal{S}^-$\;
    }
    }
    \If{$\mathcal{A} \neq \varnothing$}{
    Remove each UE $k \in \mathcal{A}$ from $\mathcal{A}$ and add it to $\mathcal{S}^-$\;
    }
\end{algorithm}

Condition (1) indicates that UE $k$ must be among the top-$q_{m}^{\rm AP}$ UEs in the updated preference list of AP $m$.
Condition (2) implies that the association ${\rm AP}_{m}-{\rm UE}_{k}$ should improve the satisfaction level of UE $k$ and should not decrease the sum of satisfaction level of all associated UEs.

Based on definition \ref{def:def2}, we propose the cluster evolution process described in \textbf{Algorithm 2}. It starts by testing the satisfaction of each UE $k$ in $\mathcal{A}$.
When UE $k$ is $\kappa_0$-satisfied (\ie $\kappa_{k}\geq \kappa_{0}$) 
, we add it to the set of satisfied UEs $\mathcal{S}^+$. Otherwise, it remains in $\mathcal{A}$ when its preference list is not empty or we add it to $\mathcal{S}^-$. 

The next steps in the Algorithm try to improve the satisfaction level of each UE $k$ remaining in $\mathcal{A}$. We start
by setting the reference index of all UEs to one ($m_{k}=1, \forall k \in \mathcal{A}$), and each UE $k$ will apply to its $m_{k}$-th preferred AP $m$ among the top-$q_{k}^{\rm UE}$ APs in its updated preference list. 
When the pair (${\rm AP}_{m}$, ${\rm UE}_{k}$) is a favorable-association pair, the association ${\rm AP}_{m}-{\rm UE}_{k}$ is set (see \textbf{Algorithm 1}). Otherwise, $m_k$ is updated as $m_{k} \leftarrow m_{k} + 1$ and UE $k$ will apply to the next $m_{k}$-th preferred AP. The cluster evolution process is repeated until there exists no favorable-association pair or $\mathcal{A}$ becomes empty. Hence, we obtain the association vector $\mathcal{C}^{\rm UE}$.

\section{Numerical results}\label{sec:result}
In this section, we evaluate the performance of our proposed algorithm, focusing specifically on its effectiveness in enhancing the satisfaction level of UEs while limiting the number of AP-UE associations. In our simulations, we consider a cell-free MIMO network consists of fixed number of $M = 50$ APs, each equipped 
with $N = 16$ antennas that jointly serve a fixed number of $K = 20$ single-antenna UEs. The APs and UEs are randomly
distributed in a 200 m by 200 m area. During $T=100$ realizations, the APs are static and UEs follow a random way-point 
mobility where the UE chooses a random direction and moves with 1 m/s. 
The UEs request random traffic data from the set $\{5, 30, 100\}$ Mb/s.
Table \ref{tab:simulation_parameters} summarizes the simulation parameters. 

\noindent
\textbf{Benchmarks.} We compare our solution with six classic benchmarks from the literature:
\begin{itemize}
    \item \textbf{Best channel (BC)} associates each UE to the AP with the highest channel gain.
    \item \textbf{Min distance (MD)} associates each UE to the closest AP based on distance.
    \item \textbf{Canonical setup (CS)} associates each UE to all APs.
    \item \textbf{Greedy combining algorithm (GCA)} forms clusters of APs for each UE by improving the minimum of QoS per UE with deactivation of certain APs \cite{Joint_AP_On_Off_and_User_Centric_Clustering_for_Energy_Efficient_Cell_Free_Massive_MIMO_Systems}.
    \item \textbf{Many-to-many without swap-matching process (\textbf{M2M w/o SMP})} associates UEs via many-to-many matching game with DA \cite{Many_to_Many_matching_user_association_scheme_in_ultra_dense_millimeter_wave_networks}.
    \item \textbf{Many-to-many with swap-matching process (\textbf{M2M w/ SMP})} is an extension of previous benchmark. After obtaining clusters, we will try to switch, between two UEs, one of their associated APs while all other associations remain unchanged. We will allow swap between two different associated APs of each pair of UEs when the substitution should not reduce the total satisfaction level of the network and should strictly increase at least $\kappa$  of one UE while $\kappa$ of the other should not decrease.
\end{itemize}

\setlength{\headsep}{0.4in}
\begin{table}[t]
    \centering
    \small
    \caption{Simulation parameters}
    \resizebox{0.91\columnwidth}{!}{
    \begin{threeparttable}
    \begin{tabular}{c c c}
    \hline
    \textbf{Notations}  & \textbf{Parameters}  & \textbf{Values \cite{Local_partial_zero_forcing_precoding_for_cell_free_massive_MIMO}} \\
       \hline
      $f$ &  Carrier frequency  &   3.5 GHz\\
      
      $B$   &  System bandwidth  &  20 MHz \\

      $T$ &  Simulation duration  &  100 \\

      $M$ &  Number of APs  &   50\\
       
     $K$   &  Number of UEs   &  20 \\

     $K_{\max}$   &  Maximum number of served UEs per AP  &   12 \\   
      
     $M_{\max}$      &  Maximum number of associated APs per UE  &  8 \\
      
     $N$   &  Number of antennas per AP  &  16 \\
       
     $R_{k}^{\rm req}$  &  Requested throughput by UE $k$  &  $\{$5, 30, 100$\}$ Mb/s  \\
      
     $P_{\rm max}$    &  Maximum power of each AP  &  200 mW \\

    $\eta$ &  Path loss exponent    &   2  \\

    $\sigma_{s}^{2}$ &   Shadowing variance   &   6  \\

    $\sigma_{n}^{2}$ &  Noise variance    &   $10^{-5}$  \\

     $D_{\mathrm{th}}$   &   Power difference threshold for APs selection   &   30 dB\tnote{1}  \\
    
     $\lambda_{\rm GCA}$ &   SE-EE ratio for GCA   &    0\tnote{2} \\
     
     $\kappa_{0}$ &  UE satisfaction level threshold   &   $\{$0.8, 0.9, 1$\}$   \\
       \hline
    \end{tabular}
    \begin{tablenotes}
        \item[1] We use the same threshold as in \cite{Joint_AP_On_Off_and_User_Centric_Clustering_for_Energy_Efficient_Cell_Free_Massive_MIMO_Systems} to form clusters of APs for each UE.
        \item[2] We choose $\lambda_{\rm GCA}=0$ to consider only minimum of SE maximization problem for the improvement process \cite{Joint_AP_On_Off_and_User_Centric_Clustering_for_Energy_Efficient_Cell_Free_Massive_MIMO_Systems}. 
    \end{tablenotes}
    \end{threeparttable}
    }
    \label{tab:simulation_parameters}
\end{table}

\noindent
\textbf{Complexity analysis.} Table \ref{tab:complexity} details the computational complexity of the different benchmarks and our proposed algorithm, which we refer to as \textbf{PA (EA-M2M)}. In particular, in the matching process of \textbf{M2M w/ SMP}, each AP sorts the list of UEs requesting association and chooses the top $K_{\rm max}$ preferred UEs. This procedure costs $\mathcal{O}(K M \log(K))$ at each iteration. In addition, the swap-matching process (SMP) in \textbf{M2M w/ SMP} requires, in the worse case, $\mathcal{O}(M_{\max}^2 K^3)$ prohibitive permutations further complexifying this approach. In contrast, our solution does not require a sorting procedure, and the cluster evolution process we propose as an alternative to SMP, limits the computational cost to $\mathcal{O}(N^{\rm EA}K)$. Therefore, \textbf{PA (EA-M2M)} significantly reduces the total execution time compared to \textbf{M2M w/ SMP}.
\setlength{\headsep}{0.21in}

\setlength{\headsep}{0.4in}
\begin{table}[t]
    \centering
    \small
    \caption{Computational complexity of compared algorithms.}
    \resizebox{\columnwidth}{!}{
    \begin{threeparttable}
    \begin{tabular}{|c|c|}
    \hline
    \textbf{User Clustering Alg.}    &  \textbf{Computational Complexity} \\
    \hline
      BC    &  $\mathcal{O}(K  M \log(M))$ \\
    \hline
      MD    &  $\mathcal{O}(K  M \log(M))$ \\
    \hline
      CS    & $\mathcal{O}(K M)$ \\
      \hline
        GCA    &  $\mathcal{O}(K M^3\log(M))$\\
    \hline
       M2M w/o SMP    & $\mathcal{O}(K  M \log(M) + K M^2 \log(K))$ \\
    \hline
        M2M w/ SMP\tnote{1}   &   $\mathcal{O}(K  M \log(M) + K M^2 \log(K) + N^{\rm SMP}M_{\rm max} K^3)$ \\

    \hline
        PA (EA-M2M)\tnote{2} &   $\mathcal{O}(K M \log(K M)  +  N^{\rm EA}\min(K_{\rm max}, M_{\rm max}) K)$ \\
    \hline
    \end{tabular}
    \begin{tablenotes}
        \item[1] Our simulations show that, in general, $N^{\rm SMP}$ does not exceed $M_{\rm max}$. 
        \item[2] $N^{\rm EA}$ denotes the number of potential tests to form favorable-association pairs given as $N^{\rm EA} = \sum_{k \in \mathcal{K}} \min(q_{k}^{\rm UE}, |\mathcal{P}_{k}^{\rm UE}|)$. It is upper-bounded by $M_{\rm max}K$.
    \end{tablenotes}
    \end{threeparttable}}
    \label{tab:complexity}
\end{table}

\begin{figure}[t]
    \centering
    \includegraphics[width=\columnwidth]{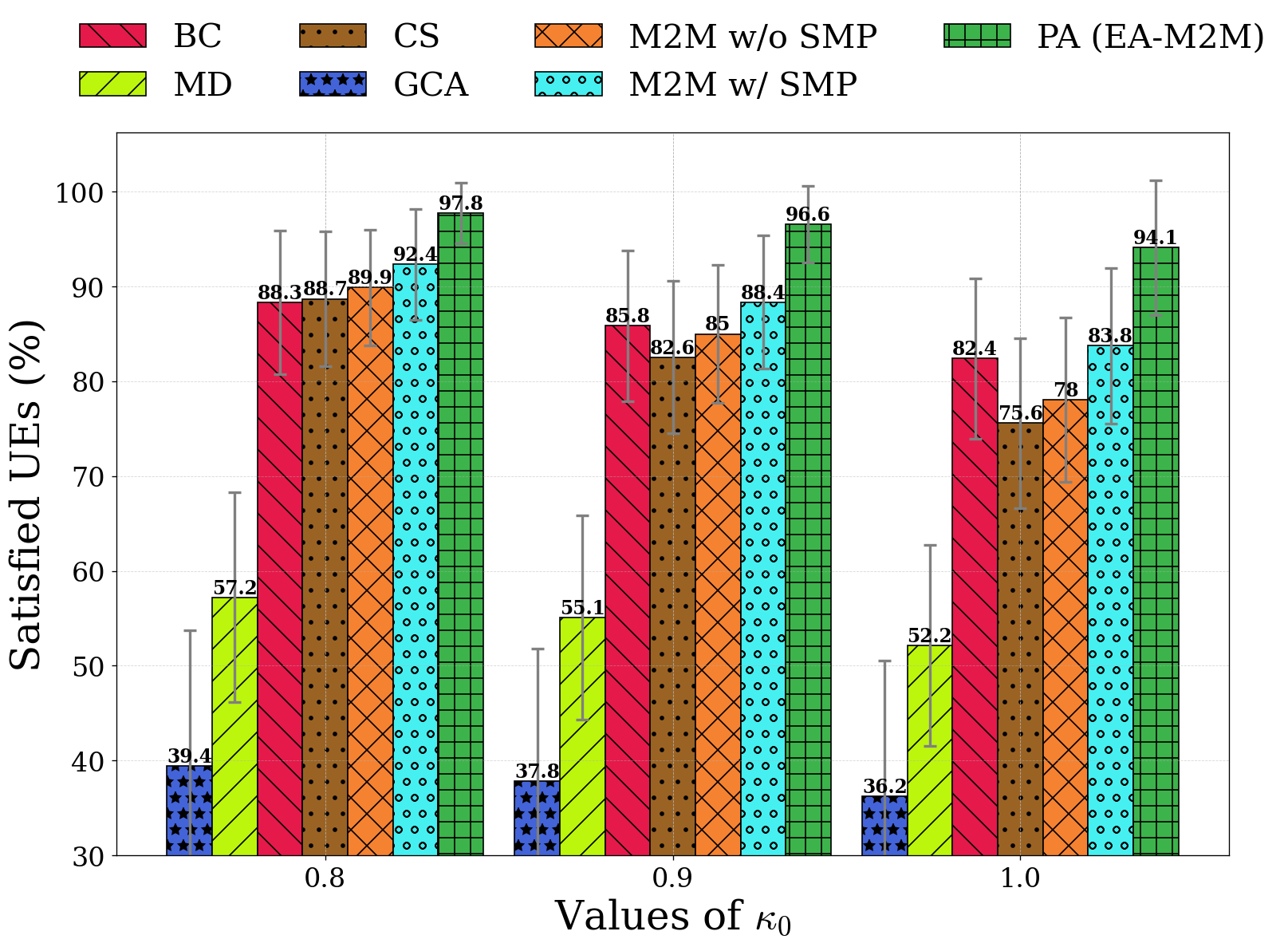}
    \caption{Percentage of $\kappa_0$-satisfied UEs in the network.}
    \label{fig:sc2_percentage_S_UEs}
\end{figure}

\noindent
\textbf{Percentage of $\kappa_0$-satisfied UEs.} This metric is exactly our objective function that we aim to maximize. 
Here, we measure how many UEs reached a $\kappa_0$-satisfaction level using our algorithm versus benchmarks. As depicted in Fig. \ref{fig:sc2_percentage_S_UEs}, our approach surpasses the benchmarks significantly, irrespective of the $\kappa_{0}$ value. Our proposed algorithm shows higher $\kappa_0$-satisfied UEs percentage that exceeds 10$\%$ in the case of fully satisfaction (\ie $\kappa_{0}=1$). Thus, we were able to improve cluster formation process with less computational cost compared to the most competitive benchmark \textbf{M2M w/ SMP}. 

\begin{figure}[t]
    \centering
    \includegraphics[width=\columnwidth]{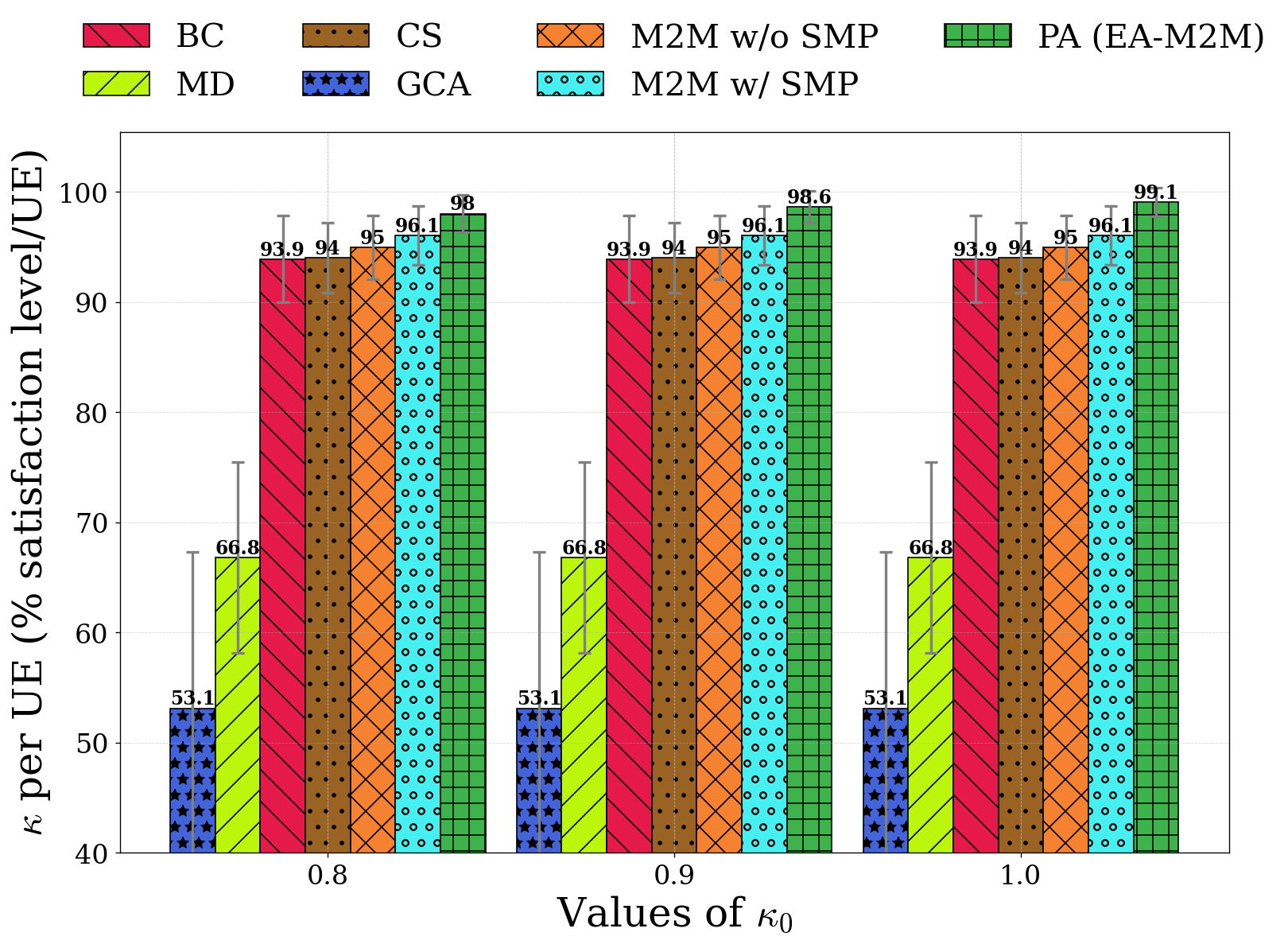}
    \caption{Percentage of satisfaction level per UE in the network.}
    \label{fig:kappa_per_UE}
\end{figure}

\begin{figure}[t]
    \centering
    \includegraphics[width=\columnwidth]{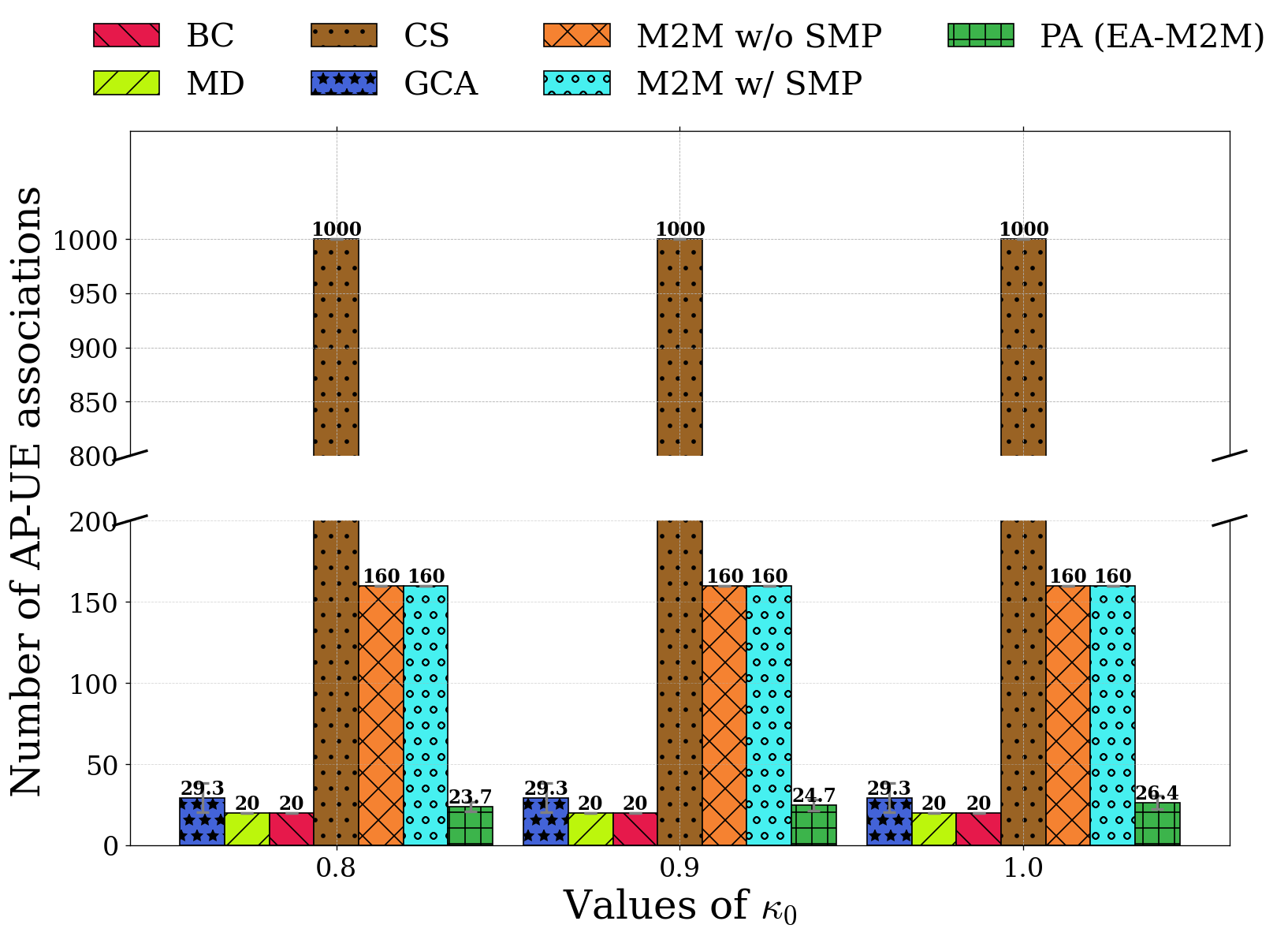}
    \caption{Number of AP-UE associations in the network.}
    \label{fig:associations_AP_UE}
\end{figure}

\noindent
\textbf{Level of satisfaction per UE.}  With this metric, we evaluate the QoS perceived by UE (\ie $\kappa$ in Eq. \eqref{eq:QoS}). As shown in Fig. \ref{fig:kappa_per_UE}, the results demonstrate that our proposed method ensures the highest satisfaction level, exceeding $98\%$ regardless of $\kappa_{0}$ value. Specifically, in scenarios where fully satisfaction is desired, our method reaches a satisfaction rate of $99.1\%$. Consequently, we guarantee that even for $5.9\%$ of UEs that do not achieve complete satisfaction (see Fig. \ref{fig:sc2_percentage_S_UEs}), they still experience a satisfactory QoS.

\noindent
\textbf{Number of  AP-UE associations.} Here, we assess the number of associations established between APs and UEs. The results presented in Fig. \ref{fig:associations_AP_UE} show that our solution significantly improves the QoS for each UE while substantially reducing the number of necessary associations between APs and UEs. Compared to \textbf{M2M w/ SMP}, which complies with the association limits of UEs and APs, our algorithm reduces by $84\%$ the number of associations required by \textbf{M2M w/ SMP}. Consequently, we significantly limit the fronthaul signaling, as each AP now interacts with a smaller number of UEs. Furthermore, it also limit the computational complexity needed for \eg channel precoding schemes.

\section{Conclusion}\label{sec:conclusion}
In this paper, we propose a matching game and devise scalable algorithms for user-centric clustering in cell-free MIMO networks. Our solution jointly takes into account the QoS requirements of UEs and limited radio resources to maximize the number of satisfied UEs in terms of requested QoS. To do so, we propose an EA-based many-to-many matching algorithm to speed the convergence time and limit the signaling overhead. Numerical results show that our proposed solution outperforms state-of-the-art benchmarks. For instance, compared to DA-based schemes, it reduces the number of associations by $84\%$, while providing up to $10\%$ improvement in the number of fully satisfied UEs. 

Future work will investigate advanced ways for defining UEs and APs preference lists to strengthen user-centric clustering process.
Moreover, our solution will include sleep mode mechanisms to improve energy efficiency of the network.

\section*{Acknowledgements}
This work was funded by the French government under the France 2030 ANR program “PEPR Networks of the Future” (ref. 22-PEFT-0003).

\bibliographystyle{ieeetr}
\bibliography{biblio}

\end{document}